\documentclass[prb,preprint]{revtex4-1} 

\usepackage{amsmath} 

\usepackage{graphicx} 

\begin{document}

\title{Improving student's problem-solving ability as well as conceptual understanding without sacrificing the physics content of a class}
\author{D. J. Webb}
\affiliation{Department of Physics, University of California, Davis, CA}
\email[]{webb@physics.ucdavis.edu}

\date{\today}

\begin{abstract}
Four sections of introductory physics for physical scientists and engineers (about 180 students each) are compared.  One section, treatment group, was organized so that students worked to learn the classical ideas connecting forces and motion over the first 6 weeks of the 10 week quarter and then used the final 4 weeks to apply those principles to algebraically complicated problems.  The other sections learned ideas at essentially the same time as calculations over the entire 10 weeks of the quarter.  The treatment group and one of the control sections were taught by the same instructor, had identical curricular materials and this instructor was blind to the comparison measure, the final exam.  After controlling for GPA as well as for incoming conceptual understanding, the treatment group was found (with greater than 99\% confidence) to perform better on the final exam than the control group taught by the same instructor and, by a similar measure, the treatment group performed significantly better than any other section.  The treatment group also had higher conceptual learning gains and so should be better prepared for later learning.
\end{abstract}

\maketitle

\section{Introduction}
A common model\cite{brown1989situated, driver1994constructing} of teaching considers the student as an apprentice learning from a master, the teacher.  This model seems particularly apt in discussions of post-secondary education.  It is essentially exactly right when discussing graduate students and still seems reasonable when the students are juniors and seniors taking major classes.  This model is not so clearly extendable to introductory classes because many of the students are not aiming at eventual expertise in physics but, instead, are aiming at other expertise such as in engineering, biology, etc.  Nevertheless, in this paper I will use this model of post-secondary education to frame the discussion of the results of a trial comparing different ways of organizing the material in a physics class.

One reason for using the apprentice-master model of teaching in this paper is that psychologists\cite{chase1973perception, chi1981categorization} have studied and compared the thinking processes of experts and of novices in various fields of study (including physics\cite{chi1981categorization}) and have developed ideas on what distinguishes these two kinds of thinker.  Although the expert knows many more facts than the novice, the more important distinction between expert and novice is that these facts are organized in the mind of the expert into patterns that are hierarchical.  For physics the lowest level of this heirarchy includes all of the specific detailed physical processes and situations that one remembers (or can construct) and the highest level includes the main overarching models that physicists use to give both qualitative and quantitative explanations of physical processes.  This expert cognitive structure allows\cite{chi1981categorization} an expert physicist to ``see'' a physical situation differently than a novice (understand the important features of the physical situation), analyze the physical situation differently (more effectively use qualitative thinking), and so solve problems more efficiently.  A student needs to learn the overarching models (i.e. construct the higher levels of the mental framework of an expert) as well as learn to connect them to specific phenomena that are new to the student.  For the intro-physics classes discussed in this paper, the overarching models include Galilean relativity, Newtonian mechanics, conservation of energy, conservation of momentum, and conservation of angular momentum.  A student initially learns the meaning of these ideas by seeing the words and equations defined and used in discussions of various physical phenomena and then strengthens their understanding through practice in qualitative and quantitative analysis of either simple or complex physical situations.  Examination of the ``difficult'' problems in almost any standard introductory-physics textbook suggests that physicists especially prize a student's ability to analyze a complicated physical situation using one or more of these overarching physical models to derive a quantitative result.  This ability is thus, apparently, one of the distinguishing abilities of an expert in the field.

The question addressed in this paper is ``Can we change the normal order of presentation of material in way that leads to increased student proficiency in the understanding that physicists prize?''.  In the last 30 years research in the field of Physics Education has shown\cite{kim2002students} that standard lectures, problem solving, and performing standard laboratory experiments do not lead to excellent student understanding of the highest level of ideas in the hierarchical mental structure of an expert physicist.  This research has also shown\cite{hake1998interactive} how to dramatically increase that understanding of the highest hierarchical level, largely by devoting classtime specifically to student-student discussions of these ideas so that students can work very directly toward building an expert conception of the ideas.  This research also generally shows\cite{hoellwarth2005direct} that, even with an increased understanding of the ideas in the highest level of the expert's mental hierarchy, students \emph{usually do not perform better} on the standard problems that physicists prize.  That is the puzzle that is addressed by this study.  Finally, though there are many excellent discussions\cite{hsu2004resource} of which types of problems a student should be practicing in order to make the best progress toward expert physicist, this paper is agnostic as to this question.  Rather, in measuring problem-solving ability, this paper uses what might be termed standard exam problems graded in the standard way.

\section{Demographics and Experiments}
UC Davis offered four lecture sections, sections I through IV, of Physics 9A (classical mechanics) in the Spring quarter of 2012.  A student in Physics 9A has three hours of lecture, one hour of discussion section, and 2.5 hours of laboratory each week.  The textbook and the laboratory experiments of all four sections were exactly the same but the lectures, discussion sections, and homework are under the control of the instructor assigned to teach that section.  The material in section I (the treatment group) was organized so that students worked to learn the classical ideas connecting forces and motion (mostly Newtonian mechanics of pointlike objects and/or rigid bodies) over the first 6 weeks of the 10 week quarter and then used the final 4 weeks to apply those principles to solve the algebraically complicated physical problems that physicist prize\cite{[{For some curricular details, including the final exam, see }]webb2012class}.  The other three sections (control groups) learned ideas at the same time as calculations over the entire 10 weeks of the quarter.  Sections I and II were both taught by the author and had identical homework problems, discussion section problems, lab problems, and lecture questions.  Sections III and IV were taught by two other instructors who each had more than 20 years of experience teaching at UC Davis, were excellent teachers, and had taught this specific course several times in the past few years.

The students had no knowledge of how any of the sections would be taught when they enrolled.  In addition, each class was completely full and almost no students could switch sections after they heard how the class would be taught, so this is a rough approximation of a randomized trial.  The average student demographics at the beginning of the quarter are shown in Table \ref{EnteringDemographics} (the errors shown are standard error in the mean).  FCI refers to the Force Concept Inventory\cite{hestenes1992force} which is a 30-question multiple-choice survey of the student's understanding of forces, motion, and their Newtonian relationship.  This survey was given to all students at the beginning of the quarter (pretest) and again (posttest) in the 8th week of the 10-week quarter at least 3 weeks after every section had finished working on the Newtonian physics of point-like objects.  We take the FCI pretest to be a measure of the initial Newtonian understandings that the students bring to the class.  The student's GPA is their UC Davis GPA at the beginning of the quarter and we take this to be a measure of their general academic ability.  ``Semesters physics'' refers to the number of semesters of previous physics classroom experience and was self-reported by the students.  Each of the sections had 15-20 more students who didn't take the FCI pretest and so are not included in the table or in later analysis.  An ANOVA analysis shows that FCI pretest distributions for the four sections were statistically indistinguishable but that section III has an incoming GPA distribution that is distinctly different from the other three lectures.  Section III also has a large number, 25, of students who were enrolled in special extra-physics-help courses (only 1 such student was in section II and 4 each in sections I and IV) and a slightly different Female fraction.  Due to these differences in the student population in section III, we leave section III out of the quantitative analysis from now on but note that our various conclusions would either be unchanged or strengthened if section III were included.

\begin{table}
\caption{\label{EnteringDemographics}Student demographics upon entering the class.  Quoted errors are standard error of the mean.}
\begin{tabular}{| p{19pt} | p{21pt} | p{58pt} | p{48pt} | p{46pt} | p{35pt} |}
\hline Lec  &   $N$  & GPA  &  FCI pretest  & Semesters physics  & Female Fract. \\ \hline
I & 157 & $2.96 \pm 0.04$ & $16.1 \pm 0.5$  & $1.9 \pm 0.1$ & 0.25 \\ \hline
II & 163  & $2.95 \pm 0.04$ & $16.3 \pm 0.5$ & $1.9 \pm 0.1$ & 0.25 \\ \hline
III & 169 & $3.18 \pm 0.04$ & $16.0 \pm 0.5$ & $1.8 \pm 0.1$ & 0.32 \\ \hline
IV & 167 & $3.04 \pm 0.04$ & $15.1 \pm 0.5$ & $1.8 \pm 0.1$ & 0.23 \\ \hline
\end{tabular}
\end{table}

We have two main outcome measures, the FCI given 8th week of the class and the final exam.  Each of the four sections took the same final exam at the same time.  This exam was written by the two instructors from sections III and IV and another instructor.  The instructor in sections I and II had no input on the final and did not see it until all instruction (including review sessions) had ended.  There were eight problems on the final exam\cite{webb2012class} and each problem was worth 20 points.  Each problem of the final exam was graded by two graders, one who graded sections I and II (mixed together) and one who graded sections III and IV (mixed together).  Each pair of graders worked together (almost all of them sat side-by-side during their grading) to normalize their grading.  This grading was supervised by the instructor from section III.  We discuss possible grading calibration issues later. 

\section{Results}
First we discuss the comparison between the two sections (I and II) that were identical except for the ordering of the topics.  These sections both had online homework using MasteringPhysics and it was easy to arrange for 60\% of the homework time to be devoted largely to conceptual and tutorial questions so that the 40\% of the homework time devoted to working on algebraically complicated problems could be moved to the final 4 weeks of the quarter in section I.  Note that \emph{both of these sections could be characterized as interactive-engagement classes}\cite{hake1998interactive}.

\begin{figure}
\includegraphics[trim=138 255 135 220, clip, width = 3.25 in]{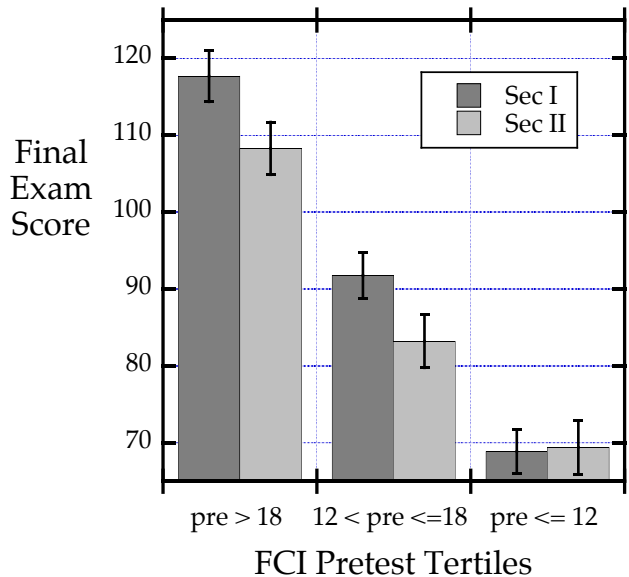}
\caption{\label{FExamByFCIpre}Final exam scores (160 pts max) comparing the treatment (section I) and control section (section II) for the class divided by tertiles determined by score on FCI at beginning of the quarter.  Error bars are standard error in the mean.}
\end{figure}

We find that the final exam score has a statistically significant dependence on three incoming measures as well as the section.  The FCI pretest score explains the largest part of the variance in the final exam scores that we control for in these two sections.  The next most important control variable is the student's UC Davis GPA upon entering the quarter followed by the number of semesters of previous physics (gender is not a significant variable as long as we control for FCI pretest).  The regression model that we fit is
\begin{eqnarray}
\label{statmodel}
FExam & = &\beta_1 (FCIpre) + \beta_2 (GPA) + \beta_2 (SemPhysExp) \nonumber  \\
&& + \beta_4 (Sect) + constant
\end{eqnarray}
where $Sect = 1$ for the treatment group and 0 for the control group.  For this model ($R^2 = 0.57$) the coefficient measuring the effect of treatment vs control is $ \beta_4 = 6.2 \pm 2.3$ $(p = 0.007)$ \cite{footnote1} so with higher than 99\% confidence we would say that students in the treatment group earned higher scores on the complicated problems that physicists ask their students to analyze.  We graphically compare treatment and control groups, as well as showing the importance of $FCIpre$, in Figure \ref{FExamByFCIpre} where we have divided the students into three equal-sized populations determined by their FCI prestest score.  The figure shows the average final exam scores in each section for each group of students and one sees that section I (treatment group) students did significantly better if they were above the lowest tertile.  If we leave the lowest FCIpre tertile out of the regression fit then the treatment group had an increase of $9.1 \pm 2.8$, 3.25 standard deviations from a null result.

\begin{figure}
\includegraphics[trim=119 255 135 220, clip, width = 3.4 in]{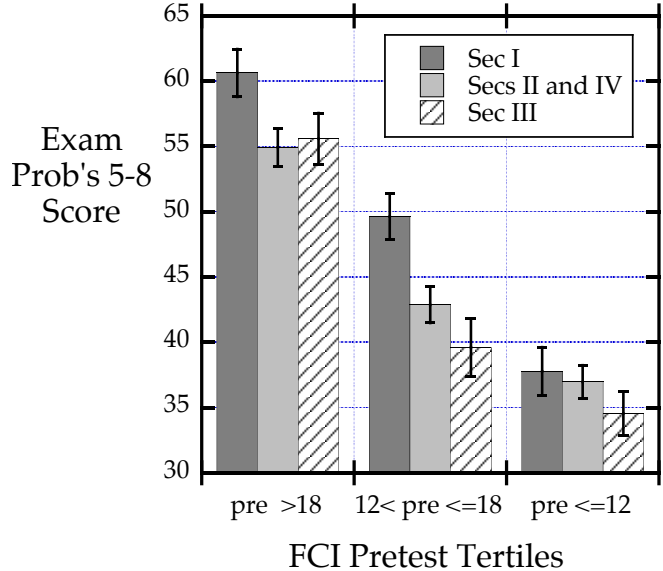}
\caption{\label{FExam58ByFCIpre}Score on four final exam problems (80 pts max) for treatment (section I) and control (sections II and IV) for the class divided by tertiles determined by score on FCI at beginning of the quarter.  Error bars are standard error in the mean.}
\end{figure}

Because the same final exam was given in all four sections, we can include section IV in the comparison and thus check against another instructor.  Unfortunately the first four problems on the final exam were identical to problems that the section IV students had seen during the quarter so we will only compare the total scores on problems 5-8 of the final exam.  We find that sections II and IV are indistinguishable in terms of their scores on problems 5-8\cite{footnote2} after we control for the three control variables we used above, so we group them together as a larger control section and compare section I with II\&IV.  Comparing I with II\&IV, we find that we must control for the same significant incoming variables and in this case the coefficient measuring the effect of treatment vs control group is $ \beta_4 = 4.7 \pm 1.2$ $(p < 0.001)$ so with higher than 99.9\% confidence we would say that students in the treatment group earned higher scores than students in the other sections on the problems that physicists expect their students to analyze.  We show this in Figure \ref{FExam58ByFCIpre} by dividing the students up, again by FCI pretest score, into three equal populations.  Notice that we have also included the results from section III in this graph to show that they are not anomalously high.  Finally, another way of characterizing the effect of the section I organization is to estimate the odds of getting a good grade on the final.  If we take A- or better as the grade we want our students to get then, at UC Davis, we expect 21\% of the students in sections II\&IV would get A- or better.  Examination of the distribution of scores shows that this will happen for these sections if we used a cutoff of 58 (out of 80) points for A- or better on these four problems.  Controlling for the same other factors as before we can use logistic regression to find that the odds of getting at least this score are $2.8 \pm 0.8$ times \emph{higher} for a student in the treatment group than for one in the control group made up of sections II\&IV.

The final outcome measure that we want to compare is the gain in conceptual understanding.  The FCI was given in the 8th week of the 10 week quarter as a posttest.  A standard way\cite{hake1998interactive} to compare classes is to calculate a class normalized gain, $g = (FCIpostavg - FCIpreavg)/(30-FCIpreavg)$ where the averages are class averages, 30 is the maximum that FCIpostavg could be, and only students who had \emph{both} FCI prestest and FCI posttest scores are included.  The results for the four sections are shown in Table \ref{FCIgains}.  As is almost always found\cite{hake1998interactive}, both sections where interactive engagement techinques are used (sections I and II) are significantly different from the standard lecture type of class, section IV.  For instance, the confidence level is greater than 99.9\% that sections I and IV have different FCI postest averages.  Figures \ref{FExamByFCIpre} and \ref{FExam58ByFCIpre} suggest that success in working the types of problems that physicist prize is much less likely if one does not enter the class having already constructed some of the highest level of the intellectual hierarchy of the expert physicist.  This learning is one result of the intro-physics class and we can quantify a treatment/control kind of comparison of the conceptual learning gains by answering the following question ``Since a student must increase their FCI score by about 7 to be fairly sure of moving up above the first tertile, how likely is this in section I vs section IV?''.  Again, we can use logistic regression (the only significant control variables are $GPA (p = 0.001)$ and $gender (p=0.026)$) to find that the odds of starting in the lowest FCI prestest group and gaining at least an additional 7 correct questions on the FCI postest are $4.2 \pm 1.7$ $(p < 0.001)$ times larger for a student who was in section I than for one who was in section IV.  Being in section I (vs. section IV) had the same size effect here as an increase in $GPA$ of 1 point.

\begin{table}
\caption{\label{FCIgains}Class-averaged values of FCI prestest and postest scores and normalized gain.  Quoted errors are standard error of the mean.}
\begin{tabular}{| p{21pt} | p{21pt} | p{55pt} | p{55pt} | p{55pt} |}
\hline Lec & $N'$ & FCI pretest  & FCI posttest & Class g (= normalized gain) \\ \hline
I & 152 &  $16.0 \pm 0.6$  & $21.3 \pm 0.5$ & $0.38 \pm 0.06$ \\ \hline
II & 151  & $16.2 \pm 0.5$ & $20.9 \pm 0.5$ & $0.34 \pm 0.06$ \\ \hline
III & 154 & $15.9 \pm 0.6$ & $19.9 \pm 0.5$ & $0.29 \pm 0.06$ \\ \hline
IV & 154 & $15.2 \pm 0.5$ & $18.4 \pm 0.5$ & $0.21 \pm 0.06$ \\ \hline
\end{tabular}
\end{table}

\section{Discussion}
This paper presented some results that are standard, in that they had been seen by many researchers, as well as the first report of a controlled study showing a new result.  The fact that the final exam scores of sections II and IV are indistinguishable from each other even though the FCI posttest scores differ at the 99.9\% confidence level is a result that is often seen in physics education research.  Learning the overarching ideas that make up the highest level of an expert's knowledge hierarchy seems to require special curricular focus on those ideas and some kind of interactive engagement by the students with the ideas (to our knowledge there is no evidence of this size conceptual learning gain in a class without explicit curricular focus on interactive engagement).  However, in the past, learning those ideas has generally not been shown to change exam scores when the exam problems are the standard problems that physicists prize.  This paper presents the first controlled parallel study whose results are consistent with significantly increased exam scores on these standard problems.  The result of this study suggests that students are better problem solvers if they have spent considerable time at the beginning of the quarter (or semester) on qualitative thinking before progressing to more computationally complicated problems later in the quarter.  We should note that there are glimpses of this general kind of result in education literature\cite{van1991overview, perry1991learning}.  Almost every physics class includes important overarching (largely qualitative) ideas and and computationally complicated applications of those ideas so the curricular materials in almost any physics class could be organized in roughly the same way as the treatment group discussed in this paper.

\begin{acknowledgments}
The author thanks Wendell Potter, Emily West, and the rest of the UC Davis PER group for useful discussion and comments.
\end{acknowledgments}


\begin{thebibliography}{5}
\bibitem{footnote1} The ``$p$-value'' is basically the probability that our result is a false-positive so $1-p$ is the confidence level.
\bibitem{footnote2}Grading studies we have done suggest systematic grading differences of at most about 2 points in the 80-point total if the graders do not talk to each other so we suggest a conservative estimate of 1 point possible systematic error for graders who are collaborating.  Errors of this size will change our numbers slightly but does not seriously affect the statistical signifigance that we quote.
\end{thebibliography}
\end{document}